\pdfoutput = 1
\documentclass[twocolumn]{article}

\usepackage{times}
\usepackage[numbers]{natbib}
\usepackage[english]{babel}
\usepackage{blindtext}
\usepackage{graphicx}
\usepackage{amsmath, amsthm, amssymb, bbm, bm}
\usepackage{enumerate}
\usepackage{float}      
\usepackage{subcaption}  
\usepackage{wrapfig}
\usepackage[margin=0cm]{caption}
\usepackage[titletoc, toc]{appendix}
\usepackage{tabularx}

\usepackage{multirow}
\usepackage{makecell}
\usepackage{placeins}  

\usepackage[x11names, usenames, dvipsnames, svgnames, table]{xcolor}
\definecolor{firebrick}{rgb}{.698,.133,.133}
\definecolor{mybluelight}{rgb}{0.9, 0.9, 1.}
\definecolor{myorangelight}{rgb}{1., 0.9, 0.9}

\usepackage[utf8]{inputenc} 
\usepackage[T1]{fontenc}    
\usepackage{url}            
\usepackage{booktabs, colortbl}       
\usepackage{amsfonts}       
\usepackage{nicefrac}       
\usepackage{microtype}      

\usepackage{csquotes}
\usepackage{latexsym}

\usepackage{pifont}
\usepackage[boxruled, vlined, linesnumbered]{algorithm2e}
\SetAlFnt{\small}
\SetAlCapFnt{\small}
\SetAlCapNameFnt{\small}
\usepackage{algorithmic}
\algsetup{linenosize=\tiny}

\let\oldnl\nl
\newcommand{\nonl}{\renewcommand{\nl}{\let\nl\oldnl}}

\usepackage{paralist}

\usepackage{xspace}
\usepackage{soul}
\usepackage{dsfont}
\usepackage{stmaryrd}
\usepackage[textwidth=15mm]{todonotes}
\usepackage{dirtytalk}
\usepackage{pbox}
\usepackage{cprotect}

\usepackage{verbatim}
\usepackage{textcomp}
\usepackage[normalem]{ulem}

\usepackage{mathtools}
\usepackage{etextools}
\usepackage[inline]{enumitem}

\usepackage[colorlinks=true,allcolors=firebrick,bookmarks=false]{hyperref}

\definecolor{darkergreen}{RGB}{21, 152, 56}
\definecolor{red2}{RGB}{252, 54, 65}
\definecolor{Gray}{gray}{0.85}
\newcolumntype{g}{>{\columncolor{Gray}}c}

\let\OLDthebibliography\thebibliography
\renewcommand\thebibliography[1]{
  \OLDthebibliography{#1}
  \setlength{\parskip}{0pt}
  \setlength{\itemsep}{0pt plus 0.3ex}
}

\theoremstyle{definition}

\DeclarePairedDelimiterX{\divx}[2]{(}{)}{%
  #1\;\delimsize\|\;#2%
}

\usepackage{style}

\makeatletter
\@namedef{ver@everyshi.sty}{}
\newcommand{\removelatexerror}{\let\@latex@error\@gobble}

\title{Leveraging Uncertainty for Deep Interpretable Classification and\\ Weakly-Supervised Segmentation of Histology Images}

\renewcommand\footnotemark{}

\author{Soufiane~Belharbi$^{1}$,
   ~Jérôme~Rony$^1$,
   ~Jose~Dolz$^2$,
  ~Ismail~Ben~Ayed$^{1}$,
  ~Luke~McCaffrey$^{3}$, and
  ~Eric~Granger$^{1}$\\
 	$^1$ LIVIA, Dept. of Systems Engineering, ÉTS, Montreal, Canada \\
 	$^2$ LIVIA, Dept. of Software and IT Engineering, ÉTS, Montreal, Canada\\
	$^3$ Goodman Cancer Research Centre, Dept. of Oncology, McGill University, Montreal, Canada\\
	{\tt\footnotesize \textcolor{black}{soufiane.belharbi.1@ens.etsmtl.ca} }
}

\newcommand{\ignore}[1]{}

\begin{document}
\maketitle\thispagestyle{fancy}

\maketitle

\begin{abstract}
Trained using only image class label, deep weakly supervised methods allow image classification and ROI segmentation for interpretability. Despite their success on natural images, they face several challenges over histology data where ROI are visually similar to background making models vulnerable to high pixel-wise false positives. These methods lack mechanisms for modeling explicitly non-discriminative regions which raises false-positive rates.
We propose novel regularization terms, which enable the model to seek both non-discriminative and discriminative regions, while discouraging unbalanced segmentations and using only image class label. Our method is composed of two networks: a localizer that yields segmentation mask, followed by a classifier. The training loss pushes the localizer to build a segmentation mask that holds most discrimiantive regions while simultaneously modeling background regions.
Comprehensive experiments\footnote{Our public code: \href{https://github.com/sbelharbi/deep-wsl-histo-min-max-uncertainty}{https://github.com/sbelharbi/deep-wsl-histo-min-max-uncertainty}.}   over two histology datasets showed the merits of our method in reducing false positives and accurately segmenting ROI.
\end{abstract}

\textbf{Keywords:} Deep Weakly-Supervised Learning, Image Classification, Semantic Segmentation, Histology Images, Interpretability

\section{Vulnerability of weakly-supervised methods to pixel-wise false positives over histology data}
Weakly-supervised learning (WSL) has seen a lot success in different applications mainly over natural images. However, recent study over histology data~\citep{rony2019weak-loc-histo-survey} showed that these methods yield high pixel-wise false positives highlighting the difficulty of transferring global labels to pixel-level in this data. This is mainly caused by two factors:
1) strong visual similarity of ROI and background in histology images making it difficult to spot ROI.
2) since WSL methods are trained to maximize their class confident, ROI segmentation is left free. Without any constraints, the model can yield an arbitrary segmentation that maximizes classification. Both factors combined lead to over-segmentation, hence high false positive. In this work, we propose to constrain ROI to be most discriminative while simultaneously modeling background regions. This increases the model awareness to the presence of background and reduces over segmentation, therefore reducing false positives.

\section{Proposed method}
\begin{figure*}[h!]
  \centering
\includegraphics[width=0.8\linewidth]{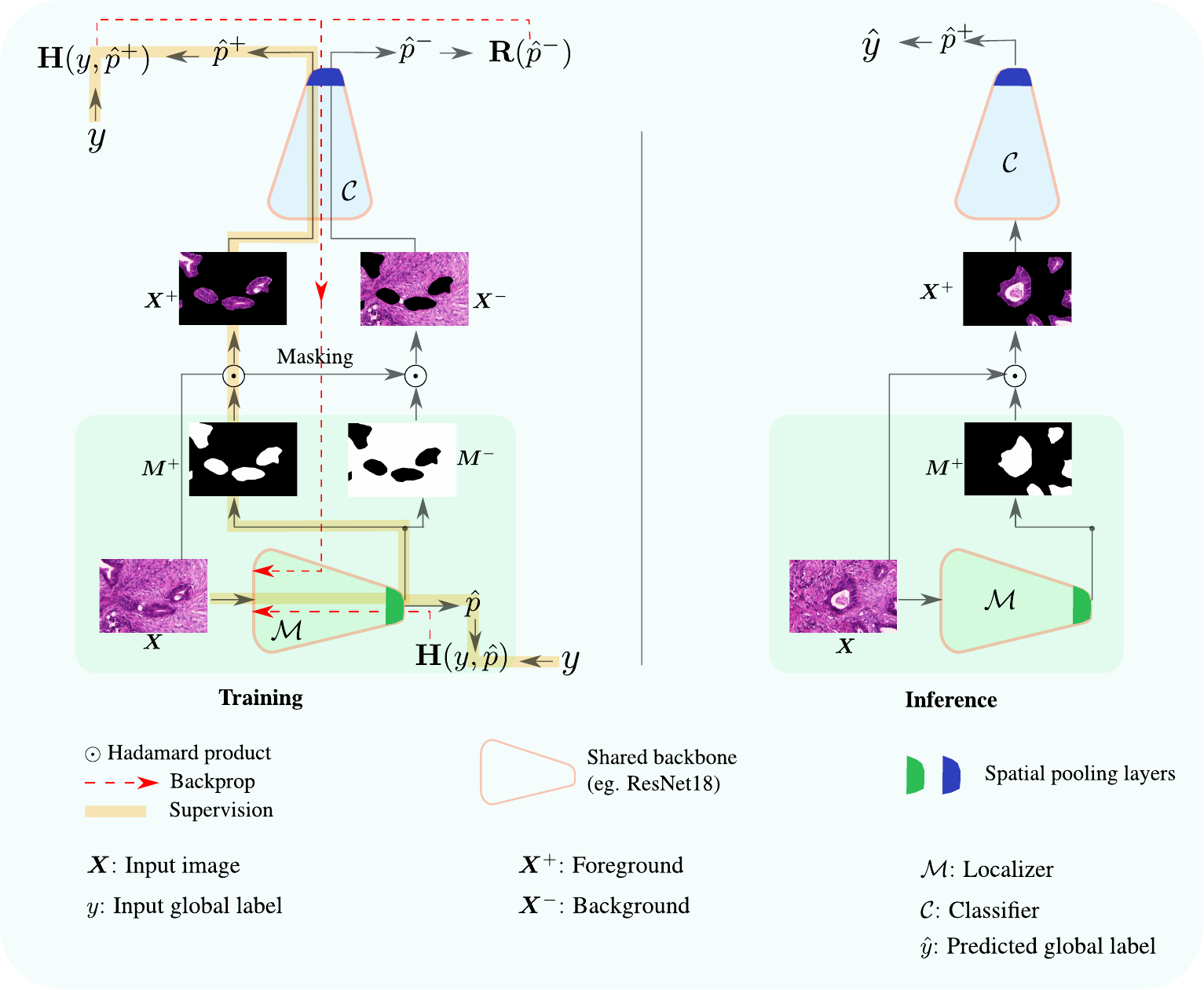}
\caption{Illustration of our approach.}
\label{fig:fig-proposal}
\end{figure*}
Fig.\ref{fig:fig-proposal} depicts training and evaluation of our proposal. It aims to simultaneously segment foreground (FG) and background (BG) regions. To do so, we constrain the classifier response $p^+$ over FG to be more certain with respect to the true class $p$ by minimizing cross-entropy $\min \mathbf{H}(p, \hat{p}^+)$. The BG is constrained to have high uncertainty with respect to the class caused by the expected absence of discriminative regions. This is achieved either by explicitly maximizing entropy $\max - \mathbf{H}(\hat{p}^-)$, or, by matching a surrogate distribution with the highest uncertainty that is uniform distribution $q$, $\min \mathbf{H}(q, \hat{p}^-)$. To avoid that one region dominates the other, we constrain the model to find the \emph{largest} FG and BG regions by imposing size constraints formulated through a log-barrier method.
\begin{equation}
\label{eq:eq-5}
\begin{aligned}
& \min_{\bm{\theta}_{\mathcal{C}}}
& & \mathbf{H}(p, \hat{p}^+) + \lambda \; \mathbf{R}(\hat{p}^-) - \frac{1}{t} \left[\log\bm{s}^+ + \log\bm{s}^- \right],
\end{aligned}
\end{equation}
where
$
\mathbf{R}(\hat{p}^-) =
   - \mathbf{H}(\hat{p}^-); $ refered as Explicit Entropy Maximization (EEM), or, $
   \mathbf{R}(\hat{p}^-) = \mathbf{H}(q, \hat{p}^-),
$ referred as Surrogate for explicit Entropy Maximization (SEM), ${\lambda}$ is a balancing positive scalar, and $t > 0$ is a parameter that determines the accuracy of the approximation of the barrier method. We define the size of each mask as: $
    \bm{s}^+ = \sum_{z \in \Omega}\bm{M}^+(z)\;, \quad \bm{s}^- = \sum_{z \in \Omega}\bm{M}^-(z)\;, $
where ${\Omega}$ is the spatial image domain. We to refer to the total size term as Absolute Size Constraints (ASC).
We use SGD to optimize Eq.\ref{eq:eq-5} which trains simultaneously the localizer and classifier in an end-to-end manner (Fig.\ref{fig:fig-proposal}).

\section{Experiments}
\begin{figure*}[h!]
  \centering
  \includegraphics[width=.9\linewidth]{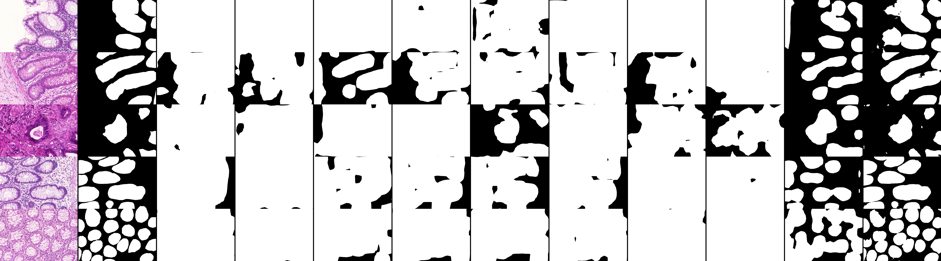} \\
  \includegraphics[width=.9\linewidth]{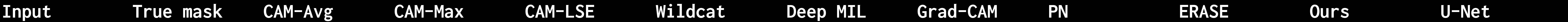}
  \caption{\textbf{GlaS dataset}: Qualitative results of the predicted binary mask for each method on several GlaS test images. Our method, referred to as \textit{Ours}, is the SEM version with the ASC regularization term. (Best visualized in color.)}
  \label{fig:glas-results-visu}
\end{figure*}

\begin{figure*}[h!]
  \centering
  \includegraphics[width=.9\linewidth]{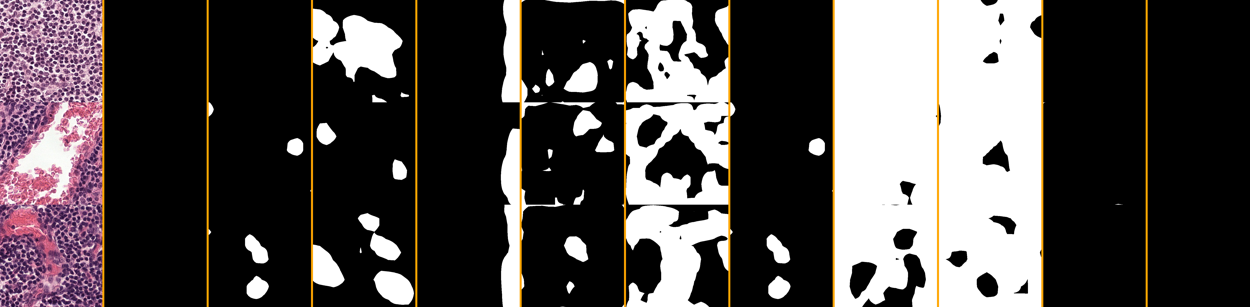} \\
  \includegraphics[width=.9\linewidth]{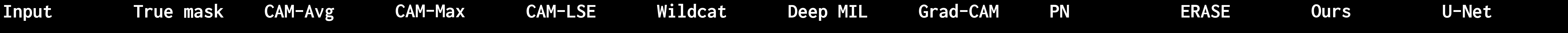}
  \caption{\textbf{Camelyon16-P512 benchmark}: Examples of mask predictions over \textbf{normal} samples from the testing set. White pixels indicate metastatic regions, while black pixels indicate normal tissue. This illustrates false positives. Note that normal samples do not contain any metastatic regions. Ours is SEM version with the ASC regularization. (Best visualized in color.)}
  \label{fig:cam16-results-visu-normal-patch}
\end{figure*}

\begin{figure*}[h!]
  \centering
  \includegraphics[width=.9\linewidth]{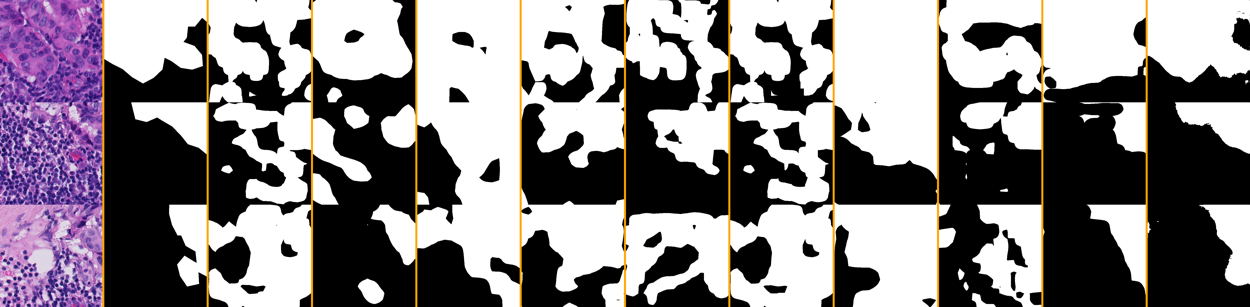} \\
  \includegraphics[width=.9\linewidth]{code-camelyon16-512-patch}
  \caption{\textbf{Camelyon16-P512 benchmark}: Examples of predicted pixel-wise masks over \textbf{metastatic} samples from the test set. White pixels indicate metastatic regions, while black pixels represent normal tissue. \textit{Ours} is the SEM version with the ASC regularization. (Best visualized in color.)}  \label{fig:cam16-results-visu-metastatic-patch}
\end{figure*}

We evaluated our method over two public histology benchmarks: GlaS\footnote{GlaS dataset: \url{https://warwick.ac.uk/fac/sci/dcs/research/tia/glascontest}} dataset for colon cancer, and patch based Camelyon16\footnote{Camelyon16 dataset: \url{https://camelyon16.grand-challenge.org}. Patches:~\citep{rony2019weak-loc-histo-survey}.} for breast cancer. For evaluation, we consider classification error for image class, and Dice index (F1) over foreground (F1$^+$) and background (F1$^-$). Quantitative results are presented in Tab.\ref{tab:tab1} which show that our method has better segmentation of ROI and low false positives. Visual results are presented in Fig\ref{fig:glas-results-visu}, \ref{fig:cam16-results-visu-normal-patch}, \ref{fig:cam16-results-visu-metastatic-patch}. The long version of this work can be found in~\citep{belharbi2022minmaxuncer}.

\begin{table*}[h!]
  \caption{Image classification and pixel-level segmentation performances on the GlaS and Camelyon16 test sets. Cl: classification. The best performance is shown in bold.
  }
  \label{tab:tab1}
  \centering
  \small
   \resizebox{.9\linewidth}{!}{
  \begin{tabular}{l|r|r|r|r|r|r}
    \toprule
    &  \multicolumn{3}{c|}{GlaS} & \multicolumn{3}{c}{Camelyon16-P512}  \\
    \midrule
    &  \multicolumn{1}{c|}{\textbf{Image level}} & \multicolumn{2}{c|}{\textbf{Pixel level}} & \multicolumn{1}{c|}{\textbf{Image level}} & \multicolumn{2}{c}{\textbf{Pixel level}} \\
    \textbf{Method} &  \multicolumn{1}{c|}{Cl. error (\%)} & \multicolumn{1}{c|}{F1$^+$ (\%)} &  \multicolumn{1}{c|}{F1$^-$ (\%)} &
    \multicolumn{1}{c|}{Cl. error (\%)} & \multicolumn{1}{c|}{F1$^+$ (\%)} &  \multicolumn{1}{c}{F1$^-$ (\%)}\\
    \midrule
    All-ones  (Lower-bound)                                                & $--$        & $66.01$      & $00.00$  & $--$        & $59.44$      & $00.00$  \\
    \midrule
    PN~\citep{kervadec2019constrained}                     & $--$        & $65.52$      & $24.08$& $--$        & $31.15$      & $37.36$  \\
    ERASE~\citep{wei2017object}                            & $7.50$       & $65.60$      & $25.01$& $8.61$      & $31.30$      & $42.48$  \\
    Max-pooling~\citep{oquab2015object}                          & $1.25$      & $66.00$      & $26.32$& $10.06$     & $48.28$      & $81.92$ \\
    CAM-LSE~\citep{PinheiroC15cvpr}                            & $1.25$      & $66.05$      & $27.93$ & $1.51$      & $64.31$      & $63.78$ \\
    Grad-CAM~\citep{selvaraju2017grad}                      & $\bm{0.00}$ & $66.30$      & $21.30$ & $2.40$      & $62.78$      & $79.05$ \\
    GAP~\citep{zhou2016learning}                         & $\bm{0.00}$ & $66.90$      & $17.88$& $2.40$      & $62.75$      & $79.05$  \\
    WILDCAT~\citep{durand2017wildcat}                        & $1.25$      & $67.21$      & $22.96$ & $\bm{1.48}$ & $62.73$      & $72.59$  \\
    Deep MIL~\citep{ilse2018attention}& $\bm{2.50}$ & $68.52$      & $41.34$ & $1.93$      & $59.01$      & $36.94$ \\
    \midrule
    Ours (EEM)                                   & $\bm{0.00}$ & $\bm{72.11}$ & $69.07$& $6.26$ & $67.98$ & $\bm{88.80}$    \\
    Ours (SEM)                                   & $\bm{0.00}$ & $71.94$ & $\bm{69.23}$& $6.95$ & $\bm{68.26}$ & $88.55$   \\
    \midrule
    U-Net~\citep{Ronneberger-unet-2015}   (Upper-bound)       & $--$        & $90.19$      & $88.52$ & $--$        & $71.11$      & $89.68$\\
    \bottomrule
  \end{tabular}
}
\end{table*}

\section{Conclusion}
Explicitly incorporating background prior in deep WSL methods has shown to reduce false positives with large margin over histology data. This was achieved by reducing over-segmentation, a common issue in WSL techniques over histology images, and in turn helps yielding accurate ROI segmentation.

\section*{Acknowledgments}
This research was supported in part by Canadian Institute of Health Research (CIHR), Natural Sciences and Engineering Research Council of Canada (NSERC) and Compute Canada.


\bibliographystyle{apalike}
\bibliography{midl-samplebibliography}

\end{document}